\documentclass[final,5p,times,twocolumn]{elsarticle}

\usepackage[utf8]{inputenc}
\usepackage[T1]{fontenc}

\graphicspath{{graphics/}}

\usepackage{amsmath}
\usepackage{amsfonts}
\usepackage{amssymb}
\usepackage{float}

\usepackage[normalem]{ulem}

\usepackage{booktabs}
\usepackage{tabularx}
\usepackage{threeparttable}
\usepackage{siunitx}

\usepackage{url}
\usepackage[colorlinks=true, citecolor=blue, linkcolor=blue, filecolor=blue,urlcolor=blue]{hyperref}

\usepackage[gen]{eurosym}

\usepackage{lineno}

\usepackage{framed} %

\usepackage{multicol} %

\usepackage{nomencl} %

\makenomenclature

\renewcommand*\nompreamble{\begin{multicols}{2}}

\renewcommand*\nompostamble{\end{multicols}}

\journal{Energy}

\begin{document}
\begin{frontmatter}

\title{The Benefits of Cooperation in a Highly Renewable European Electricity Network}

\author[fias]{D.P.~Schlachtberger\corref{cor1}}
\ead{schlachtberger@fias.uni-frankfurt.de}
\author[fias]{T.~Brown}
\author[fias]{S.~Schramm}
\author[aarh]{M.~Greiner}

\cortext[cor1]{Corresponding author}

\address[fias]{Frankfurt Institute for Advanced Studies, 60438 Frankfurt am Main, Germany}
\address[aarh]{Department of Engineering, Aarhus University, 8000 Aarhus C, Denmark}

\begin{abstract}
To reach ambitious European CO$_2$ emission reduction targets, most
scenarios of future European electricity systems rely on
large shares of wind and solar photovoltaic power generation. We interpolate between two concepts for balancing
the variability of these renewable sources: balancing at continental
scales using the transmission grid and balancing locally
with storage. This interpolation is done by systematically restricting
transmission capacities from the optimum level to zero.
We run techno-economic cost optimizations for the capacity investment
and dispatch of wind, solar, hydroelectricity, natural
gas power generation and transmission, as well as storage options such
as pumped-hydro, battery, and hydrogen storage. The
simulations assume a 95\% CO$_2$ emission reduction compared to 1990,
and are run over a full historical year of weather and
electricity demand for 30 European countries. In the cost-optimal system
with high levels of transmission expansion, energy
generation is dominated by wind (65\%) and hydro (15\%), with average
system costs comparable to today's system. Restricting
transmission shifts the balance in favour of solar and storage, driving
up costs by a third. As the restriction is relaxed, 85\% of the cost benefits of the optimal grid expansion
can be captured already with only 44\% of the
transmission volume.

\end{abstract}


\end{frontmatter}

\begin{table*}[!t]

\begin{framed}

\nomenclature[01]{$n$}{nodes (countries)}
\nomenclature[02]{$t$}{hours of the year}
\nomenclature[03]{$s$}{generation and storage technologies}
\nomenclature[04]{$\ell$}{inter-connectors}%

\nomenclature[05]{$c_{n,s}$}{fixed annualised generation and storage costs}
\nomenclature[06]{$c_\ell$}{fixed annualised line costs}
\nomenclature[07]{$o_{n,s}$}{variable generation costs}
\nomenclature[08]{$\lambda_{n,t}$}{locational marginal price}
\nomenclature[09]{$\mu_{LV/\mathrm{CO}_{2}}$}{KKT multipliers / shadow prices}%

\nomenclature[10]{$e_{s}$}{specific CO$_{2}$ emissions}%

\nomenclature[11]{$d_{n,t}$}{demand}
\nomenclature[12]{$g_{n,s,t}$}{generation and storage dispatch}
\nomenclature[13]{$\bar{g}_{n,s,t}$}{availability per unit of capacity}
\nomenclature[14]{$G_{n,s}$}{generation and storage capacity}
\nomenclature[15]{$G_{n,s}^{max}$}{maximum installable capacity}%

\nomenclature[16]{$soc_{n,s,t}$}{storage state-of-charge}
\nomenclature[17]{$\eta_{s}$}{generation and storage efficiency}
\nomenclature[18]{$E_{n,s}$}{storage energy capacity}
\nomenclature[19]{$h_{s,max}$}{(dis-)charge time at max. power}%

\nomenclature[20]{$f_{\ell,t}$}{power flow}
\nomenclature[21]{$F_{\ell}$}{transmission capacity}
\nomenclature[22]{$K_{n\ell}$}{incidence matrix}
\nomenclature[23]{$l_\ell$}{length of transmission line}
\nomenclature[24]{$LV$}{line volume} %
\nomenclature[25]{$f_{n-1}$}{n-1 security factor}%

\nomenclature[26]{$c_{CP}$}{capital cost of AC-DC converter pair}%

\nomenclature[27]{EU}{European Union}%

\nomenclature[27]{KKT}{Karush-Kuhn-Tucker} 
\nomenclature[27]{LMP}{locational marginal price}
\nomenclature[27]{O\&M}{operation and maintenance}%

\nomenclature[27]{PV}{solar photovoltaic}
\nomenclature[27]{OCGT}{open-cycle gas turbines}
\nomenclature[27]{PHS}{pumped hydro storage}
\nomenclature[27]{H$_2$}{molecular hydrogen}%

\nomenclature[27]{HVDC}{high-voltage direct current}
\nomenclature[27]{HVAC}{high-voltage alternating current}
\nomenclature[27]{NTC}{net transfer capacity}%

\printnomenclature

\end{framed}

\end{table*}

\section{Introduction}

The European Council has set the goal to reduce CO$_2$ emissions in
the European Union by between 80\% and 95\% in 2050 compared to their
1990 values \cite{roadm2050}. Most European countries will rely on
renewable energy sources to reach this goal. Although the majority of
renewable energy comes from hydroelectricity today, the renewable
sources with the greatest expansion potential are wind and solar
energy.

The strong weather-dependent variations of wind and solar generation
present a challenge to the balancing of production and demand in the
electricity system. These variations have particular spatial scales
(wind speeds have a correlation length of several hundreds of
kilometres) and temporal scales (both solar and wind have daily
variations, but also seasonal patterns and synoptic-scale variations
of multiple days as large weather systems pass). The countries of
Europe are small enough that the wind and solar generation inside each
country is highly correlated. This means that if each country has to
balance its own electricity generation, it must be able to deal
with the extreme highs and lows of wind and solar generation by
itself. Because exploitable hydroelectricity sites are limited and
geographically very unevenly distributed, and backup generation from
fossil fuel plants is restricted by the CO$_2$ cap, the rest of the
balancing must come from storage solutions or, in part, from demand
side flexibility.  The need to invest in storage, on top of generation
assets, tends to make these electricity systems expensive
\cite{Czisch,Scholz,Wagner2016}. %

The alternative is to balance the fluctuations of wind and solar in
space with inter-connecting transmission between countries, rather
than in time with storage. These solutions require networks on the
continental scale in order to smooth over the varying feed-in caused
by synoptic-scale weather systems. Since the costs of the required
transmission infrastructure are significantly lower than either
storage or generation assets, these systems tend to be more
cost-effective than storage-based systems
\cite{Czisch,Scholz,Schaber,Schaber2,rolando2014,Hagspiel,Brown}. However, they require large
expansions of transmission capacity that seem implausible in the face
of low public acceptance for overhead power lines \cite{Battaglinietal2012}.

Previous studies have explored the extreme points of this dichotomy
between networks and storage \cite{Czisch,Scholz}. The main innovation
of this work is to interpolate smoothly between a continent-scale
network-dominated system and a locally-balanced storage-dominated
system by continuously varying the allowed volume of transmission
inter-connectors, from zero up to unlimited interconnection. This
reveals non-linearities in the behaviour of system costs as
transmission is expanded. It is shown that most of the benefits of grid
expansion can be achieved with only a moderate expansion, which is an
important conclusions for policy-makers confronting public acceptance
issues arising from new transmission projects.

This study falls into a class of studies of the future European
electricity system where load and generation are aggregated at the
country level. It follows the work of \cite{Czisch,Scholz} by taking a
cost-optimal linear programming approach to investment while
restricting CO$_2$ emissions, but unlike these studies it explores
parameter sweeps in the space of solutions to reveal non-linear effects
as constraints are continuously tightened. The parameter space
approach can also be found in the more stylised studies of
\cite{Heide2010,sarah,rolando2014,Sensitivity}, where the effects of
different shares of wind and solar energy on backup generation and
transmission needs are explored.  In contrast to those parametric
studies, the results here incorporate realistic modelling of
hydroelectric resources, given their importance as an existing source
of low-carbon backup flexibility, other sources of storage, a fully
heterogeneous allocation of wind and solar capacities to countries
within geographic potentials, and a focus on CO$_2$ reduction rather
than increasing available renewable energy. The study \cite{OptHet}
introduces a cost-optimal, heterogeneous allocation of wind and solar
capacities around Europe, finding a sizeable reduction in total costs
compared to homogeneous distributions, but does not incorporate hydro,
storage or CO$_2$ reduction in the modelling. A genetic algorithm is
used in \cite{Bussar201440} to optimise capacities and dispatch over
three years in Europe, the Middle East and North Africa to compute
storage requirements, but does not incorporate reservoir or
run-of-river hydroelectricity.

Other classes of studies of the optimal European electricity system
model the transmission networks of each country in more spatial detail
\cite{Egerer,Hagspiel}, but because of computational limits, they can
only consider a small number of representative weather conditions,
which cannot capture the full spatio-temporal correlations across the
continent. In contrast, this study considers a full year of weather
situations.  Other studies only optimise the transmission expansion
while fixing the generation fleet
\cite{Egerer,Schaber,Schaber2,Brown}; this has the disadvantage that
the optimisation cannot weigh up whether it is better to build
renewables far from load centres and transport the energy, or build
renewables closer to demand. In this study generation and transmission
capacities are optimised jointly.

The study presented here only considers the electricity
sector. Coupling electricity to the transport, heating, cooling and
industrial energy sectors may provide additional sources of
flexibility that can help to integrate variable renewables. Studies of
sector coupling have in the past either considered single countries
(see e.g. Denmark \cite{mathiesen2014smart,Lund201296}, Germany \cite{Henning20141003,IEESWV,Quaschning}, and Ireland
\cite{Deane2012303}) or considered the whole of Europe but without
optimising international cooperation \cite{Connolly20161634}. In an
upcoming paper we will consider a full optimisation of electricity,
heating and transport in the European context. Preliminary results
\cite{Brown2016} show that the coordinated charging of battery electric
vehicles and thermal energy storage can replace much of the need for
stationary electricity storage when transmission expansion is
restricted.

A further distinction of the model presented here is that the
modelling framework uses free software \cite{PyPSA} and all the
model-specific code, input data, and output data will be available
online \cite{zenodo}, in order to further the transparency and
reproducibility of the results.

In this paper first results from our study are analysed, starting with
an introduction to the mathematical model in Section \ref{sec:model}
and the data inputs in Section \ref{sec:data}. In Section
\ref{sec:res} the results are presented from the point of view of
total costs, energy production and the interplay between spatial
distribution and temporal variations. Finally in Section
\ref{sec:diss} the results are discussed and compared to other studies
in the literature, before conclusions are drawn in Section
\ref{sec:concl}.

\section{Methods: Model}
\label{sec:model}

In this study a future, highly renewable European electricity network
is modelled. The capacities and dispatch of renewable energy
generators are optimised within each country according to their
geographical and weather-dependent potentials, with the goal of
reaching ambitious CO$_2$ reduction targets. Examples of the output
capacities can be found in Figure \ref{fig:maps_Opt}, while a sample
dispatch for a single country is shown in Figure \ref{fig:dispatch}.

\subsection{Objective function}

The model is formulated as a techno-economic linear optimization
problem that minimizes the total annual system costs.
If nodes are labelled by $n$, generation and storage technologies at the node
by $s$, hours of the year by $t$ and inter-connectors by $\ell$, then
the total annual system cost consists of fixed annualised costs
$c_{n,s}$ for generation and storage capacity $G_{n,s}$, fixed
annualised costs $c_\ell$ for transmission capacity $F_{\ell}$ and
variable costs $o_{n,s}$ for generation and storage dispatch
$g_{n,s,t}$. Costs are not associated with the flow $f_{\ell,t}$ on inter-connector $\ell$ in hour $t$. The objective function is then
\begin{equation}
  \min_{G_{n,s},F_\ell,g_{n,s,t},f_{\ell,t}} \left( \sum_{n,s} c_{n,s} G_{n,s} + \sum_{\ell} c_{\ell} F_{\ell} + \sum_{n,s,t} o_{n,s} g_{n,s,t} \right) \label{eq:objective}
\end{equation}

The optimization has to satisfy a number of constraints described in the following.

\subsection{Power balance constraints}

To ensure a stable operation of the network, energy demand and generation have to match in every hour in each node. If the inelastic demand at node $n$ and time $t$ is given by $d_{n,t}$ then
\begin{equation}
  \sum_{s} g_{n,s,t} - d_{n,t} = \sum_{\ell} K_{n\ell} f_{\ell,t} \hspace{1cm} \leftrightarrow \hspace{0.5cm} \lambda_{n,t} \hspace{0.5cm} \forall\, n,t \label{eq:balance}
\end{equation}
where $K_{n\ell}$ is the incidence matrix of the network.

The Karush-Kuhn-Tucker (KKT) multiplier $\lambda_{n,t}$ associated with the constraint indicates the
marginal price of supplying additional demand at node $n$ in hour $t$,
also known as the Locational Marginal Price (LMP). The value of
$\lambda_{n,t}$ at the optimal point is an output of the optimisation.
Background on the use of KKT duality in electricity markets can be
found in \cite{Schweppeetal1988,Biggar}.

\subsection{Generator constraints}

The dispatch of conventional generators is constrained by the capacity $G_{n,s}$
\begin{equation}
  0 \leq g_{n,s,t} \leq G_{n,s} \hspace{1cm} \forall\, n,s,t
\end{equation}

The maximum producible energy per hour in each installed unit of the renewable generators depends on the current weather conditions, which is expressed as an availability $\bar{g}_{n,s,t}$ per unit of its capacity:
\begin{equation}\label{eq:availability}
 0 \leq  g_{n,s,t} \leq \bar{g}_{n,s,t} G_{n,s} \hspace{1cm} \forall\, n,s,t
\end{equation}
Note that excess energy can always be curtailed, e.g., by pitch regulation of wind turbines or spillage in hydro power plants.
Only reservoir hydro power plants can delay the dispatch of the natural inflow to some extent by utilizing the storage reservoir.

The installed capacity itself is also subject to optimisation, with a maximum limit $G_{n,s}^{max}$ set by the geographic potential:
\begin{equation}\label{eq:instpot}
 0 \leq  G_{n,s} \leq  G_{n,s}^{max} \hspace{1cm} \forall\, n,s
\end{equation}

The capacity $G_{n,s}$ and the final dispatch $g_{n,s,t}$ of each generator are determined in the optimisation such that they respect the physical constraints above, while minimising the total costs summed in the objective function \eqref{eq:objective}.

\subsection{Storage operation}

The state-of-charge $soc_{n,s,t}$ of all storage units has to be consistent with the charging and discharging in each hour, and less than the storage capacity
\begin{align}
  soc_{n,s,t} & = soc_{n,s,t-1} + \eta_{1} g_{n,s,t,\mathit{charge}} - \eta_{2}^{-1} g_{n,s,t,\mathit{discharge}} \nonumber \\
  &\qquad + g_{n,s,t,\mathit{inflow}} - g_{n,s,t,\mathit{spillage}} , \\
0 \leq  soc_{n,s,t} & \leq h_{s,max} \cdot G_{n,s}   \hspace{1cm} \forall\, n,s,t
\end{align}
The efficiencies $\eta_1, \eta_2$ determine the losses during charging and discharging, respectively. %
These losses also imply that the storage is only charged when there is oversupply of power available in the system, and discharged when the generators can not produce enough power and the import options are not sufficient.
The state-of-charge is limited by the energy capacity $E_{n,s} = h_{s,max} \cdot G_{n,s}$. Here, $h_{s,max}$ is the fixed amount of time in which the storage unit can be fully charged or discharged at maximum power. %
In this model, reservoir hydroelectricity storages can be charged by natural inflow of water, which has to be spilled should the reservoir already be full in a given hour.

The state-of-charge is assumed to be cyclic, i.e., it is required to be equal in the first and the last hour of the simulation: $soc_{n,s,t=0} = soc_{n,s,t=T}$. This is reasonable when modelling a full year, due to the yearly periodicity of demand and seasonal generation patterns, and allows efficient usage of the storage at the beginning of the modelled time range.

\subsection{Inter-connecting transmission}

The transmission lines between countries are treated as a transport
model with controllable dispatch (a coupled source and sink),
constrained by energy conservation at each node. This is considered
to be a justifiable approximation because many of the international connections are already
controllable point-to-point high-voltage direct current (HVDC) connections, such as those undersea
(like France-Britain), those over land (like the Spain-France INELFE project) or those in the planning phase (like the HVDC link planned between Germany and Belgium),
while the flow on borders with only high-voltage alternating current (HVAC) connections are
being increasingly controlled by phase-shifting transformers (like the
German-Dutch, German-Polish and German-Czech borders). This also
follows the way that interconnectors are handled in market clearing with Net Transfer
Capacities (NTCs) on many borders.

The absolute flows on these transmission lines cannot exceed the line capacities due to thermal limits:
\begin{equation}
  |f_{\ell,t}| \leq F_{\ell} \hspace{1cm} \forall\,\ell,t
\end{equation}
The line capacities $F_{\ell}$ can be expanded by the model if it is cost-effective to do so. To satisfy n-1 security requirements, a safety margin of 33\% of the installed capacity can be used \cite{DENAII,Brown}. This
can be %
emulated a posteriori by increasing the optimized NTCs by a factor of $f_{n-1}=(1-\mathit{margin})^{-1}=1.5$. %

The lengths of the interconnecting transmission lines $l_\ell$ are set by the
distance between the geographical mid-points of each country, so that
some of the transmission within each country is also reflected in the
optimisation. A factor of 25\% is added to the line lengths to account
for the fact that transmission lines cannot be placed as the crow
flies due to land use restriction. It is assumed that there is
sufficient grid capacity within each country to redistribute power as
necessary. This assumption is driven by the decision to focus on
long-distance interconnecting transmission, which enables the 
leveraging of continental smoothing effects of interest here, but the assumption may
not always be reasonable, given that there are already North--South
grid bottlenecks in Germany, for example. However, many more spatially-detailed
studies \cite{Brown,TYNDP2016} show that total transmission costs are
typically small compared to the total generation investment cost. The
cost impact only becomes significant if the internal transmission
lines cannot be built because of missing public acceptance, which then
drives up generation costs if the best sites cannot be exploited. This
trade-off is the subject of a forthcoming paper \cite{hoersch2017arxiv}.

The sum of transmission line capacities multiplied by their lengths is
restricted by a cap $\mathrm{CAP}_{LV}$ which is varied in different simulations:
\begin{equation}
  \sum_{\ell} l_\ell \cdot F_{\ell} \leq  \mathrm{CAP}_{LV} \hspace{1cm} \leftrightarrow \hspace{0.5cm} \mu_{LV} \label{eq:lvcap}
\end{equation}
Line capacities are weighted by their lengths because the length
increases both the cost and public acceptance problems of the
transmission lines. The cap, measured in MWkm, was raised from zero to
the point where the constraint was no longer binding. The
KKT multiplier, or shadow price, $\mu_{LV}$
indicates the marginal value of an increase in line volume $LV$ to the
system; it can also be interpreted as the cost per MWkm necessary for
the optimal solution to have the transmission volume
$\textrm{CAP}_{LV}$ if the constraint \eqref{eq:lvcap} is deactivated.

\subsection{CO$_2$ emission constraints}

CO$_2$ emissions are also limited by a cap $\mathrm{CAP}_{CO_{2}}$, implemented using the
specific emissions $e_{s}$ in CO$_2$-tonne-per-MWh of the fuel of generator type $s$
and the efficiency $\eta_{s}$ of the generator:
\begin{equation}
  \sum_{n,s,t} \frac{1}{\eta_{s}} g_{n,s,t}\cdot e_{s} \leq  \mathrm{CAP}_{CO_{2}} \hspace{1cm} \leftrightarrow \hspace{0.5cm} \mu_{CO_{2}} \label{eq:co2cap}
\end{equation}
The KKT multiplier $\mu_{CO_{2}}$ indicates the carbon dioxide price
necessary to obtain this reduction in emissions in an unconstrained market.

\subsection{Software implementation}

The model was implemented in the PyPSA \cite{PyPSA} modelling
framework and was optimised using the logarithmic barrier algorithm of
the Gurobi \cite{gurobi} solver software. Using this algorithm the
model typically solves in $1-2$ hours per scenario on the local
compute node (which has multiple Intel Xeon CPU cores rated at 2.3~GHz
and 128~GB of RAM). This provides solutions whose accuracy can be
measured by the closeness of the duality gap, which in all simulations
was at most $2\cdot 10^{-6}$ of the total objective value.

\section{Methods: Data}
\label{sec:data}

The data underlying this model is presented in this section.

\subsection{Network Topology}
Following \cite{rolando2014}, the model consists of 30 nodes with one node per country of the EU-28, excluding Cyprus and Malta, but including Norway, Switzerland, Serbia, and Bosnia and Herzegovina.
The nodes are connected with the topology of the already existing international transmission lines (see Fig.~\ref{fig:maps_Opt} for the topology).

\subsection{Time series}

The model is run for a full year with hourly resolution. The year 2011 was chosen because it is the earliest available year with full availability of the input data. %
The hourly electricity demand in each country is based on \cite{Heide2010,entsoe_load}.
The onshore wind, offshore wind, and solar photovoltaic (PV) power generation are based on historic weather data with hourly temporal and a $40 \times 40~\text{km}^2$ spatial resolution over Europe using a similar method as described in \cite{Heide2011,Andresen20151074}. This method first converts weather data to potential power generation time series in each raster cell and then aggregates the results on country level, weighted by a spatial distribution of generators. This sets the availability $\bar{g}_{n,s,t}$ per unit of capacity of the renewable generation (cf.~\eqref{eq:availability}).

\subsection{Capacity layouts}

The capacity layouts of the three renewable resources in each country were set proportional to the usable area and the potential full load hours per raster cell, such that sites with higher average power production are preferred and the average full load hours are relatively high.
The available area was restricted by the following constraints:
Onshore wind and PV can only be built in areas with certain land use types defined by the CORINE database \cite{corine2006}, following the selection reported by \cite{Scholz}.
Therefore, wind farms are not placed in urban areas and solar panels are not built in forests, for example.
Offshore wind sites were restricted to a maximum water depth of 50 m.
Additionally, all nature reserves and restricted areas defined in the Natura database \cite{natura2000} are excluded. Note that this source did not include such data for non-EU-28 countries, and therefore possible restrictions in these countries are not considered in this study.

\begin{figure*}%
\centering
\includegraphics[width=\linewidth]{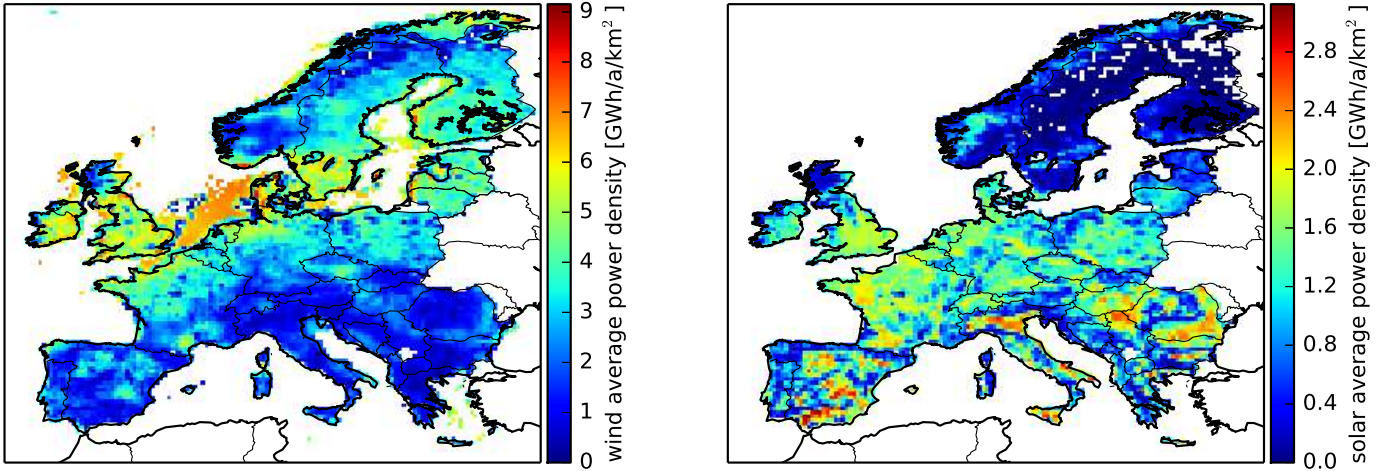}
\caption{Potential average power density for installable wind (left) and solar (right) generation per $\sim 40 \times 40 \mathrm{km}^{2}$ raster cell over Europe, once various land use restrictions have been taken into account. Raster cells with zero values and areas outside the considered regions are white.
}
\label{fig:EUlayout}
\end{figure*}

Fig.~\ref{fig:EUlayout} indicates that the spread of potential full load hours of onshore wind is large in large countries.
The average full load hours are an important factor in determining the spatial distribution of installed capacity. If the distribution of full load hours is very inhomogeneous, the average depends strongly on the region size
after determining the capacity layout as described above.
In order to get a better estimate of the economic efficiency of wind turbines in each country and to allow a fair comparison between countries of different sizes, we split the onshore wind layout of the ten largest countries into up to four equal area parts. The spatial distribution of the new parts in each country is defined by similar full load hours.
This procedure increases the spatial resolution of the onshore wind generation by adding independent classes of generators with different time series and average full load hours to the single node of a country. Their optimized capacities and produced energies are later aggregated again on country level for analysis.

\subsection{Geographic potential}

The geographic installation potential $G^{max}_{n,s}$ is also based on these layouts. It is assumed that in each country each renewable capacity can be extended proportional to this layout only until the installation density reaches a threshold somewhere. The maximum installation density of both onshore and offshore wind power is assumed to be 10 MW/km$^{2}$ \cite{Scholz}. Additionally, it is assumed that only a 20\% fraction of the already restricted area is available for installation of wind generators due to competing land use and likely public acceptance issues. This leads to an effective threshold of 2 MW/km$^{2}$. For the same reasons, only up to 1\% of the area can be used for solar PV panels with a nominal capacity of 145 MW/km$^{2}$ \cite{Andresen20151074}. This results in a potential installable energy density per raster cell that is shown in Fig.~\ref{fig:EUlayout}. Most of the best wind conditions are located along the North Sea, the Baltic Sea, and the Aegean Sea, while the highest solar potentials are in Spain and south-east Europe.

\subsection{Hydroelectricity}

Reservoir hydro and run-of-river power plants can convert water inflow into electricity. Both generation types are assumed to remain at their currently installed capacity, i.e., are not expanded, due to environmental concerns, which defines a conservative lower bound.
Data on country-specific installed hydro power capacities is provided by \cite{kies2016,pfluger2011}, but does not distinguish between reservoir and run-of-river types. Therefore, the power capacities were split proportional to the run-of-river share per country published by \cite{ENTSOEinstalledcapas}. This source did not report run-of-river shares for some countries,
in which case the estimated shares collected within the Restore2050 project \cite{kies2016} were taken instead.
All energy storage capacities from \cite{kies2016} are attributed to the reservoir hydro power plants, for a total of 207.6 TWh.

The inflow time series per country are based on \cite{kies2016}, where daily river run-off data \cite{dee2011} was weighted by the respective geographic height and normalized to match yearly generation data.
The total inflow in each country was split into reservoir and run-of-river inflow, proportional to the shares of installed power capacity.

\subsection{Non-renewable generators}

The renewable generator portfolio in each country can be complemented by conventional backup generators.
Their global annual energy generation is limited by a strict European CO$_2$ emission limit corresponding to a reduction of
95\% compared to 1990, but they can be dispatched independent of weather conditions and therefore help to provide sufficient power even in the most extreme hours.
The conventional backup system is represented here by open-cycle gas turbines (OCGT), following the assumptions of \cite{schroeder2013}, due to their high flexibility and load-following capabilities, and relatively low capital costs, such that they require few full load hours per year to be economically feasible.

\subsection{Storage}

Fluctuating generation can also be mitigated to some extent via temporal
shifting in storage units. In this study, three types of storage technologies
are considered: pumped hydro storage (PHS), central batteries, and hydrogen
(H$_2$) storage with electrolyzers, fuel cells, and above-ground steel tanks \cite{budischak2013}.
All three storage types are modelled with equal nominal charging and discharging power capacities, respectively.
Storage and dispatch efficiencies may differ, however, as listed in Table~\ref{tab:costsassumptions}.
The storage energy capacities %
are assumed to be proportional to the power capacities %
such that a storage unit can be fully charged or discharged at maximum power in a fixed amount of time $h_{s,max}$.
These simplifications are done to limit computational effort and partly due to lack of detailed publicly available data.
The storage energy standing loss over time is not explicitly included in the model.

The already existing pumped hydro storage capacities reported by \cite{kies2016}
are assumed to remain in use, but without additional extension potential. This assumption is slightly conservative, but further PHS potentials in Europe are estimated to be small due to environmental concerns.
Capital costs for existing units are neglected due to the
 long technical lifetimes and site-specific investment contingencies  of hydro power plants.
PHS units are typically designed to provide short term load shifting within a
day. Assuming an appropriate dimensioning of storage energy capacity, the latter
is set via $h_{\mathrm{PHS},max} = 6\, \text{h}$.

The installation potential of batteries and H$_2$ storage is not constrained.
Central batteries have high round-trip efficiencies but relatively high storage
losses over time. Therefore, they are best suited for short-term storage, and
are modelled here with $h_{\mathrm{battery},max}=6\, \text{h}$.
H$_2$ storage can provide a long-term storage option due to relatively low
efficiencies but low losses over time. A relatively large energy capacity is
chosen with $h_{\mathrm{H}_{2},max}=168\, \text{h}$, i.e., one week.

\subsection{Cost assumptions}

All cost assumptions are summarized in Table~\ref{tab:costsassumptions}. The given overnight capital costs were converted to net present costs with a discount rate of $7\%$ %
over the economic lifetime.

No expansions of hydro reservoir, run-of-river, and pumped hydro storage capacities are considered in this study and the already existing facilities are considered amortized. Therefore, only their fixed operation and maintenance (O\&M) costs are taken into account when calculating the total system cost.

The transmission investment per line $\ell$ is calculated as:
$(400 \euro/\si{kW}/\si{km} \cdot 1.25 l_\ell + c_{CP}) f_{n-1}$
with converter pair costs $c_{CP} = 150000 \euro/\si{MW}$, and n-1 security factor $f_{n-1} = 1.5$. %
The fixed operation and maintenance costs for transmission lines are 2\% of the (length-dependent) investment cost.

\begin{table*}%
\begin{threeparttable}
\caption{Input parameters based on 2030 value estimates from \cite{schroeder2013} unless stated otherwise.}
\label{tab:costsassumptions}
\begin{tabularx}{\textwidth}{lrrrrrrr}
\toprule

Technology & \multicolumn{1}{l}{investment} & \multicolumn{1}{l}{fixed O\&M} & \multicolumn{1}{l}{marginal} & \multicolumn{1}{l}{lifetime} & \multicolumn{1}{l}{efficiency} & \multicolumn{1}{l}{capital cost per} & \multicolumn{1}{l}{$h_{max}$} \\
 & \multicolumn{1}{l}{(\euro/kW)} & \multicolumn{1}{l}{cost} & \multicolumn{1}{l}{cost} & \multicolumn{1}{l}{(years)} & \multicolumn{1}{l}{(fraction)} & \multicolumn{1}{l}{energy storage} & \multicolumn{1}{l}{(h)} \\
 & \multicolumn{1}{l}{} & \multicolumn{1}{l}{(\euro/kW/year)} & \multicolumn{1}{l}{(\euro/MWh)} & \multicolumn{1}{l}{} & \multicolumn{1}{l}{} & \multicolumn{1}{l}{(\euro/kWh)} &  \\

\midrule

onshore wind & 1182 & 35 & 0.015\tnote{a} & 25 & 1 &  &  \\
offshore wind & 2506 & 80 & 0.02\tnote{a} & 25 & 1 &  &  \\
solar PV & 600 & 25 & 0.01\tnote{a} & 25 & 1 &  &  \\
OCGT\tnote{b} & 400 & 15 & 58.4\tnote{c} & 30 & 0.39 &  &  \\
hydrogen storage\tnote{d} & 555 & 9.2 & 0 & 20 & \multicolumn{1}{r}{$0.75\cdot 0.58$\tnote{e}} & 8.4 & 168 \\
central battery (LiTi)\tnote{d} & 310 & 9.3 & 0 & 20 & \multicolumn{1}{r}{$0.9 \cdot 0.9$\tnote{e}} & 144.6 & 6 \\
transmission\tnote{f} & {400 \euro /MWkm} & 2\% & 0 & 40 & 1 &  &  \\
PHS & 2000\tnote{g} & 20 & 0 & 80 & 0.75 & N/A\tnote{g} & 6 \\
hydro reservoir & 2000\tnote{g} & 20 & 0 & 80 & 0.9 & N/A\tnote{g} & fixed\tnote{h} \\
run-of-river & 3000\tnote{g} & 60 & 0 & 80 & 0.9 &  &  \\
\bottomrule
\end{tabularx}

\begin{tablenotes}
\item [a] The order of curtailment is determined by assuming small marginal costs for renewables.
\item [b] Open-cycle gas turbines have a CO$_2$ emission intensity of 0.19 t/MW$_{th}$.
\item [c] This includes fuel costs of 21.6 \euro/MWh$_{th}$.
\item [d] Budischak et al. \cite{budischak2013}.
\item [e] The storage round-trip efficiency consists of charging and discharging efficiencies $\eta_1 \cdot \eta_2$.
\item [f] Hagspiel et al. \cite{Hagspiel}.
\item [g] The installed facilities are not expanded in this model and are considered to be amortized. %
\item [h] Determined by size of existing energy storage \cite{ENTSOEinstalledcapas,kies2016}.

\end{tablenotes}
\end{threeparttable}
\end{table*}

\section{Results}
\label{sec:res}
\subsection{Total costs as function of line volume constraints}

\begin{table}
\centering
\caption{Optimized average system costs in [\euro/MWh] for the allowed total interconnecting line volume of the zero, today's, compromise, and optimal grid scenarios. Also given are the overall average local marginal prices (LMP) and the total line volume.}
\begin{tabularx}{\columnwidth}{lrrrr}
  \toprule
 {\bf Scenario}  & {\bf Zero} & {\bf Today} &  {\bf Comp.} & {\bf Opt.} \\
  \midrule
Line vol. [TWkm]   &  0.0   &  31.25  &  125.0 &  285.70 \\
\midrule
battery storage    &     9.9 &     8.5 &     4.5 &     1.7 \\
hydrogen storage   &     8.1 &     5.4 &     3.4 &     3.1 \\
gas                &     4.6 &     4.2 &     4.1 &     4.5 \\
solar              &    26.1 &    21.8 &    14.7 &     9.4 \\
onshore wind       &    22.3 &    20.3 &    23.4 &    28.6 \\
offshore wind      &    10.8 &    12.0 &    11.4 &     7.5 \\
transmission lines &     0.0 &     1.0 &     3.6 &     7.6 \\
PHS                &     0.3 &     0.3 &     0.3 &     0.3 \\
run-of-river       &     1.4 &     1.4 &     1.4 &     1.4 \\
reservoir hydro    &     0.8 &     0.8 &     0.8 &     0.8 \\
\midrule
Total cost         &    84.1 &    75.7 &    67.5 &    64.8 \\
\midrule
Average LMP       &   116.5 &   107.5 &    97.4 &    90.1 \\
\bottomrule
\end{tabularx}
\label{tab:results_costs}
\end{table}

\begin{figure}%
\centering
\includegraphics[width=\linewidth]{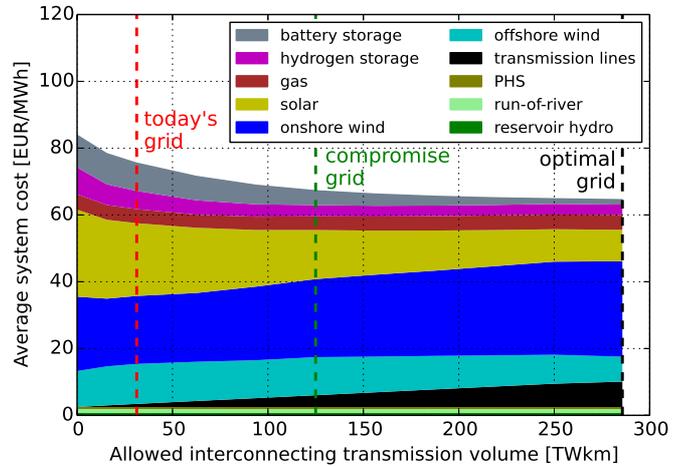} %
\caption{%
Optimized average total system costs per unit of generated energy in
\euro{}/MWh as function of the allowed total line volume between zero
transmission and the cost optimal volume. The total costs are divided into costs for the modelled components battery storage (grey), H$_2$ storage (magenta), gas (red), solar (yellow), onshore wind (blue), offshore wind (cyan), and transmission lines (black), top to bottom. The dashed vertical lines mark the transmission line volumes of today's grid (red), the compromise grid (green) at four times today's volume, and the economically optimal grid (black). Larger allowed line volumes cannot add value.}
\label{fig:costs_LV}
\end{figure}

Fig.~\ref{fig:costs_LV} shows the composition of the average cost for all investment and operation of the optimized highly-renewable European system as a function of the allowed volume of transmission lines (set by the cap in equation \eqref{eq:lvcap} and measured in MWkm). In this graphic transmission costs are set assuming the costs for overhead lines are used. The results are also given in Table \ref{tab:results_costs}.

Note first that the development of the costs is highly non-linear as
transmission volume is reduced. As the volume is restricted from the
optimal point, the costs barely increase; at this point the solution
space is very flat, i.e., costs are insensitive to restricting
transmission expansion. Only when the transmission volume is
restricted to a few multiples of today's grid\footnote{Today's grid is taken to be the Net Transfer Capacities (NTCs) between countries, multiplied by the line lengths defined in the model.}, the costs start to
increase very steeply, driven by bigger investments in storage
technologies and solar power.

In the economically optimal scenario, the total average cost for a
highly renewable power system is 64.8 \euro{}/MWh.
For comparison the cost of today's European system can be estimated from the currently installed net generation capacities and yearly energy generation for 2013 \cite{ENTSOEYSAR2013} combined with technology cost assumptions from the same source \cite{schroeder2013} as between 52 and $61\euro/\si{MWh}$, depending on whether the decommissioning and waste disposal costs for nuclear power are included. This indicates that highly renewable scenarios can have system costs that are comparable to today's system cost. In these estimates, potential CO$_2$ emission prices are neglected, which would predominantly increase costs in the conventional system.
Although transmission investments contribute
only 12\% to the total cost, the optimum line volume is 286~TWkm,
roughly 9 times higher than today's NTCs of
31~TWkm.

Such a large grid extension seems to be infeasible due to social acceptance issues \cite{Battaglinietal2012}.
On the other hand, restricting transmission requires more storage to deal with variability, driving up the costs by up to 30\% compared to the economic optimum.
However, the cost development between these two extremes is not linear: most of the increase occurs at small allowed line volumes, while the cost curve is quite flat closer to the optimum.
This allows a compromise grid of four times today's NTCs to lock in 85\% of the cost reduction of the optimally extended grid compared to the case without transmission grid, giving an average cost of 79.9~\euro/MWh.
Today's line volume, optimally distributed, would lock in just 43\% of the benefit.

If the composition of costs is examined, wind power installations contribute around 32 to 36 \euro{}/MWh to the average system cost, relatively independent of the allowed transmission volume.
For high line volumes, this is the dominant part of the cost and reflects the fact that most of the energy is generated by wind.
However, only the cost share of onshore wind increases roughly linearly from 22.3 to $28.6 \euro/\si{MWh}$ with interconnection volume while the offshore wind share decreases accordingly.
The non-linear reduction of the total system costs is due to the decreasing contributions of solar, batteries, and H$_2$ storage. All three show a similar behaviour of a strong decrease at low transmission volumes that levels off towards the optimal grid volume.
Solar costs account for a third of total cost at first but add only $9.4\euro/\si{MWh}$ in the optimal case.
The cost share of batteries and H$_2$ storage is relatively small with respectively up to 9.9 and 8.1$\euro/\si{MWh}$, and in the economically optimal scenario almost no batteries are installed.

Onshore wind is the only system component whose cost share increases with transmission volume. %
Weather patterns over Europe are typically correlated over synoptic spatial scales of roughly 1000 km, such that there are usually a few independent wind regions at all times. European-wide transmission therefore allows direct power balancing between these regions, which increases the efficient use of the wind generators.
In contrast, solar PV profits much less from this smoothing effect because one of its dominant variabilities is due to the day-night cycle that affects all of Europe almost at the same time. Transmission also allows the sites with higher capacity factors to be exploited more fully, which further increases efficiency.

However, if transmission is strongly limited, most power balancing must be done locally with the help of storage or, if available, dispatchable sources. Wind power is less effective in this case because the wind pattern typically vary on a time scale of 3 to 10 days. Shifting the demand over such periods requires large amounts of energy storage capacity, e.g., from long-term hydrogen storage.
It is therefore cost effective to install a larger share of solar, where the energy often has to be stored only between day and night.

Additionally, increasing the transmission volume allows to share temporarily unused and long-term storage between countries, which makes them more cost efficient and can help to reduce the installation demand. It also enables better access to the existing dispatchable and pumped hydro power facilities that are mostly located in Scandinavia and the Alps.

Offshore wind has the least volatile generation and can therefore provide relatively continuous power also to neighbouring countries and has the least storage needs. This is beneficial as long as line volumes are restricted, but due to the high investment costs of offshore wind, it is gradually replaced by a combination of less capital-intensive onshore wind generation and smoothing via an extended grid. %
For very small line volumes, offshore exports are also limited by grid congestion, which leads to a slight reduction of installations.

The costs for the three types of hydro power is determined by the fixed operation and maintenance cost of their assumed installation capacities and was not subject to the optimization. They are therefore constant at 2.5 \euro/MWh throughout all scenarios.

\subsection{Energy mix}

\begin{table}
\centering
\caption{Optimized annual energy generation in [\%] of the annual energy demand for the allowed total interconnecting line volume of the zero, today's, compromise, and optimal grid scenarios.}
\begin{tabularx}{\columnwidth}{lrrrr}
\toprule
 {\bf Scenario}  & {\bf Zero} & {\bf Today} &  {\bf Comp.} & {\bf Opt.} \\
  \midrule
Line vol. [TWkm] &  0.0   &  31.25  &  125.0 &  285.70 \\
\midrule
gas             &     5.1 &     5.1 &     5.1 &     5.1 \\
solar           &    40.1 &    34.9 &    24.6 &    16.2 \\
onshore wind    &    37.2 &    37.4 &    46.6 &    59.0 \\
offshore wind   &    13.8 &    15.6 &    13.9 &     8.6 \\
run-of-river    &     4.9 &     5.0 &     5.0 &     5.0 \\
reservoir hydro &     9.5 &    10.0 &    10.0 &    10.0 \\
\midrule
Total energy    &   110.6 &   108.0 &   105.2 &   104.0 \\
\bottomrule
\end{tabularx}
\label{tab:results_energy}
\end{table}

\begin{figure}%
\centering
\includegraphics[width=\linewidth]{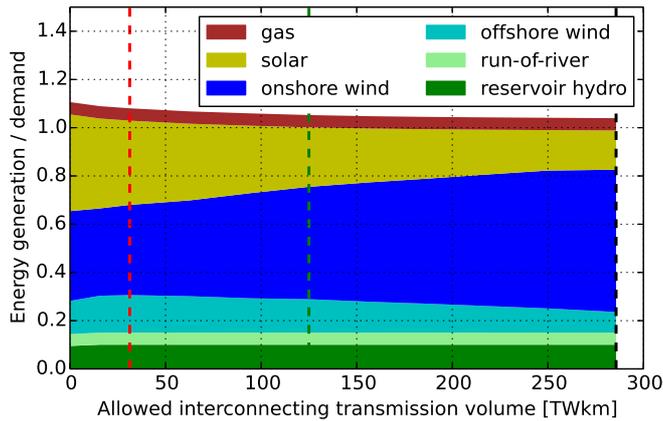}%
\caption{Optimized annual dispatched energy generation in units of the annual energy demand as function of the allowed total line volume. The total energy is divided into generation from the modelled components gas (red), solar (yellow), onshore wind (blue), offshore wind (cyan), run-of-river (light green), and reservoir hydro (green), top to bottom. As in Fig.~\ref{fig:costs_LV}, the dashed vertical lines mark the transmission line volumes of today's grid (red), the compromise grid (green) at four times today's volume, and the economically optimal grid (black).
Energy generation above the demand is caused by losses from storage use.
The amount of curtailed energy is not shown.}
\label{fig:energy_LV}
\end{figure}

The composition of energy generation per year in units of the total demand 3152 TWh/a as a function of transmission line volume is shown in Fig.~\ref{fig:energy_LV} and Table~\ref{tab:results_energy}.
The energy mix is dominated by wind which contributes 46\% to 65\% of the generation, mostly from onshore generators. Its share increases with the line volume as large scale wind variations can be smoothed better by a larger grid. This is consistent with the trends already indicated by the cost analysis. The large contribution from wind relative to the cost share shows an efficient utilization of the installed capacity.
The amount of energy contributed by solar PV generation is relatively high  with 40\% of the demand as long as transmission is strongly restricted, but decreases to only 16\% with optimal grid extension.

There is a clear correlation between the share of solar generation and the excess production required to compensate losses from storage. The latter is indicated by values above 1 in Fig.~\ref{fig:energy_LV}. It decreases with allowed interconnection from 11\% to 4\% of the demand.
This indicates that systems with a lack of transmission require a more diverse energy mix with relatively high shares of solar generation and storage use, while additional transmission increases both the economic and energy efficiency.

The constant contribution from run-of-river and reservoir hydro of 15\% of the demand is equal to the inflow and indicates a negligible need for spillage.
The energy from gas power plants is limited by CO$_{2}$ emission constraints to 5.1\% of the demand in all cases.

In hours with potential excess variable renewable generation that can neither be consumed or stored, this energy is curtailed. The total curtailed energy is reduced from 11\% of demand (340~TWh/a) with no transmission to 9\% of demand (286~TWh/a) with optimal transmission.
There is an inherent ambiguity in the order of curtailment between generators with equal, i.e., zero marginal costs. This ambiguity was lifted by introducing different small costs for each generator type (see Tab.~\ref{tab:costsassumptions}). Offshore wind generation has the highest marginal cost and is curtailed first, followed if necessary by onshore wind, solar, and finally run-of-river generation. Therefore, the contributions from run-of-river and solar are as high as possible.
This introduces a bias in the mix of produced energy.
However, this bias does not affect the installed capacities and system costs, but only redistributes the effective full load hours. In practice, a suitable reimbursement mechanism for curtailing could easily limit the economic consequences. The curtailment order can be justified by practical manageability considerations: it is easier to curtail the same amount of power in a few large off-shore wind parks than a large number of decentral solar panels.

\subsection{Spatial distribution of infrastructure}

\begin{figure}%
\centering

\includegraphics[width=\linewidth]{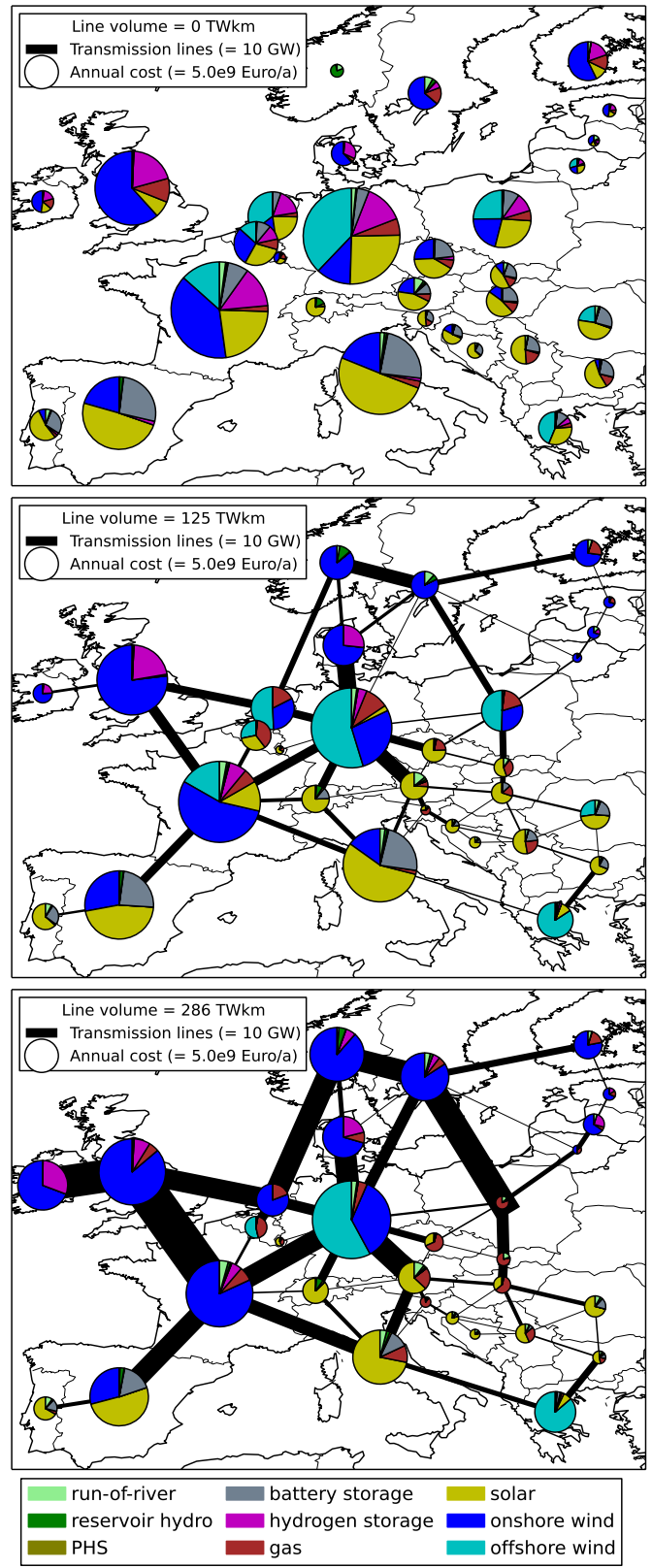}

\caption{Distributions of the cost composition per country as pie charts for the case of zero interconnecting transmission (top), compromise grid (middle), and economically optimal transmission (bottom).
The color code is the same as in Fig.~\ref{fig:costs_LV}.
The area of the circles is proportional to the total costs per country.
The modelled international transmission lines are shown as %
black lines with width proportional to their optimized net transfer capacity.
}
\label{fig:maps_Opt}
\end{figure}

\begin{figure*}%
\centering
\includegraphics[width=0.9\linewidth]{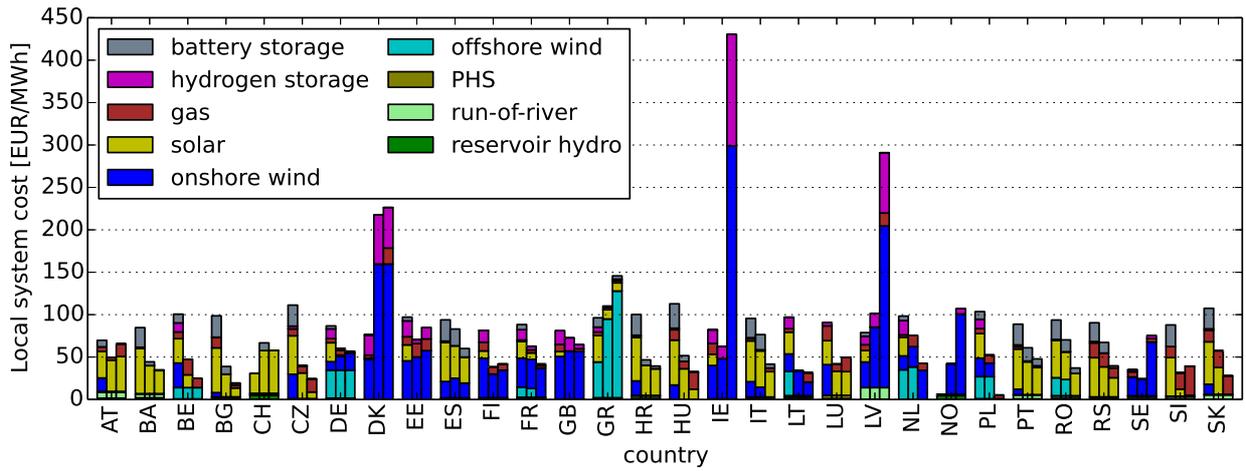}
\caption{Generation and storage costs normalised by annual demand for each country. The left, middle, and right bars are for the zero, compromise, and optimal transmission scenarios, respectively.
Energy generation can be above the local demand due to storage losses and exports.
The color code is the same as in Fig.~\ref{fig:energy_LV}.
The 30 modelled countries are ordered alphabetically by their ISO2 code.
}
\label{fig:costs_ct}
\end{figure*}

\begin{figure*}%
\centering
\includegraphics[width=0.9\linewidth]{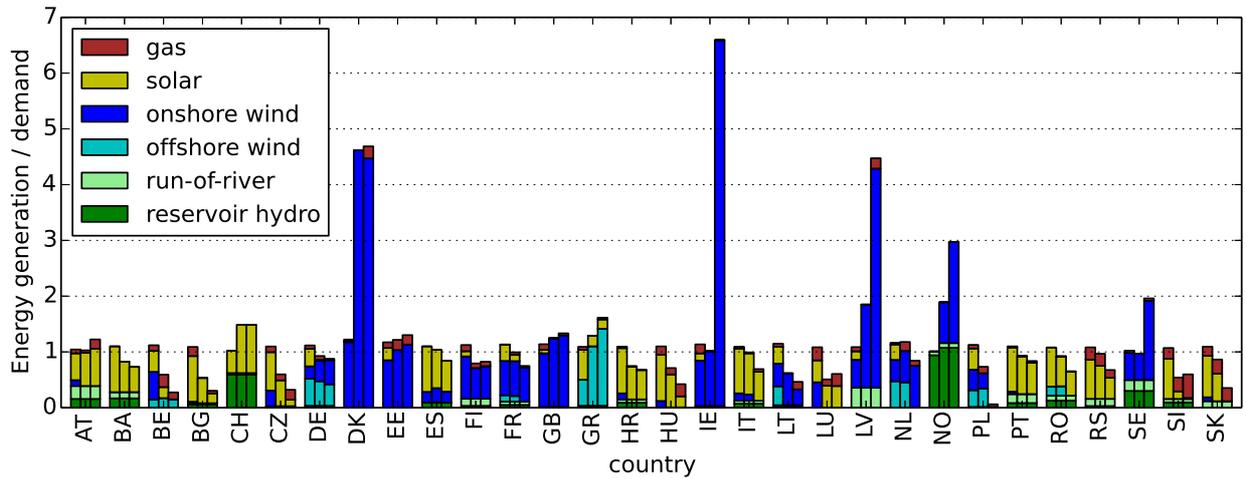}
\caption{Energy generation in units of demand per country split by generator type. The left, middle, and right bars are for the zero, compromise, and optimal transmission scenarios, respectively.
Energy generation can be above the local demand due to storage losses and exports.
The color code is the same as in Fig.~\ref{fig:energy_LV}.
}
\label{fig:installed_capacity}
\end{figure*}

The spatial distributions of the optimized annual costs for generation, storage and transmission are shown in Fig.~\ref{fig:maps_Opt} for the three scenarios of allowed transmission line volume of no transmission, the compromise grid expansion, and the economically optimal grid. The same data is graphed in Fig.~\ref{fig:costs_ct} for ease of comparison, normalised to the average load in each country. In Fig.~\ref{fig:installed_capacity} the total energy generation is plotted, normalised by each country's demand.

\begin{figure*}%
\centering
\includegraphics[width=0.9\linewidth]{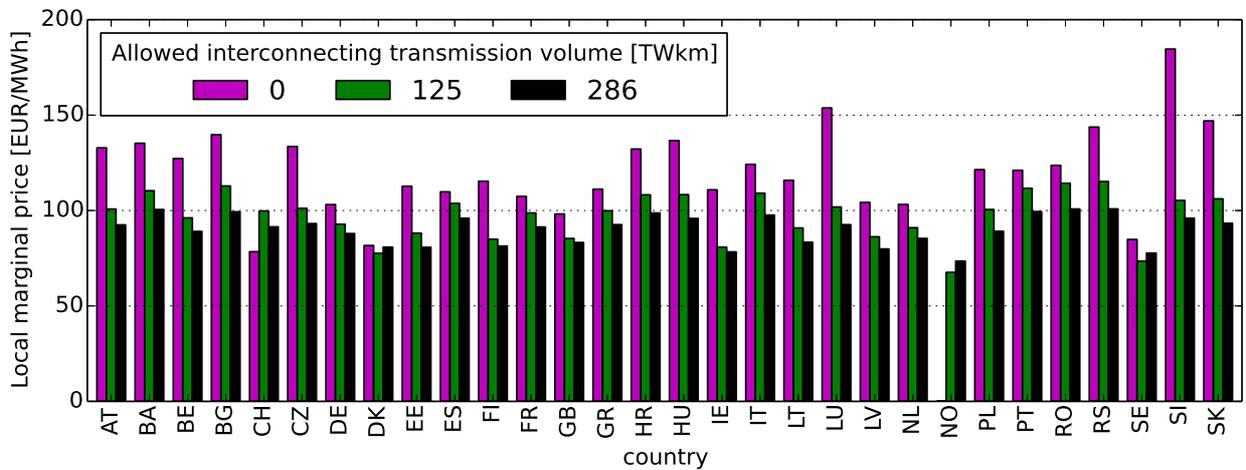}
\caption{Average marginal prices at each node for the zero (purple), compromise (green) and optimal (black) transmission scenarios.
}
\label{fig:lmp}
\end{figure*}

First we consider the case without transmission grid. Although this case is unrealistic given that countries are already inter-connected today, it provides a useful reference point to assess the benefits of cooperation.
Without transmission grid,
there is a very diverse mix of energy sources where almost every country has
wind, solar, and gas generation, with the highest shares of solar in southern Europe.
This technological diversification is cost-effective in the absence of transmission because the different resource characteristics on different time scales can be
used for some country internal balancing, as discussed, e.g., by \cite{rolando2014}. %
However, this still requires significant amounts of storage in most countries, with a relatively homogeneous share of around 20\% of the costs, similar to its share of the total costs as shown in Fig.~\ref{fig:costs_LV}. The mix of H$_2$ and battery storage especially in the central European countries reflects the composition of generation technologies: countries with high shares of wind power have high shares of longer-term H$_2$ storage, while solar dominated countries usually have larger, or exclusively, shorter-term battery capacities.
The existing pumped hydro storage units have properties similar to those of batteries and also allow a slightly increased share of solar generation.

The dispatchable reservoir hydro power is highly beneficial where available by providing a flexible source of energy that can be stored until needed. This helps to balance the mismatch between fluctuating renewable generation and demand and can reduce costs considerably. Countries with large hydro capacities like Norway and Switzerland have to build relatively little other infrastructure.

Offshore wind power often has high capacity factors and relatively little volatility but high installation costs. Ten countries also utilize their offshore wind resources in this scenario. In Greece, the Netherlands, and Germany the investment shares are above 25\%. It is not built in Great Britain, Ireland, and Denmark which have some of the best offshore wind efficiencies, but also very good onshore wind conditions and large installation potentials. There, long term H$_2$ storage and the differing characteristics of solar provide more benefits than additional offshore wind.

In the scenario with the compromise grid expansion,
there is already some splitting of resource utilization between different
regions in Europe in order to better exploit the available potentials.
Due to relatively weak solar irradiation in northern Europe, solar generation is no longer built there, which also largely removes the benefits of battery storage in this region.
Instead, only a few countries like Denmark, Great Britain, and Norway with very good wind potentials increase the size of their onshore wind installations and export the excess energy. It is then cost efficient for the other countries to install similar amounts or even less onshore wind capacity than in the scenario without transmission.
The grid infrastructure is built predominantly between large and wind
dominated countries in north-western Europe, where it helps to increase the generation efficiency further via synoptic-scale spatial smoothing of wind fluctuations.
This also reduces the need for H$_2$ storage in all countries with wind installations.
Only Denmark and Great Britain increase their H$_2$ storage capacities slightly, which suggests that their exports can be supported by congestion management through storage.

In south-eastern Europe, the solar and battery installations are almost the same as in the previous scenario, but the introduction of relatively small transmission capacities allows to replace the local wind installations with imports from nearby regions. Exceptions are Greece and Romania that have access to good offshore wind resources, that are now utilized even more in Greece and used for exports.

The technology mix especially in the larger countries in central
and south-western Europe is still heterogeneous due to compromises between resource efficiency and transmission constraints. This is also indicated by the concentration of balancing power generation from gas power plant in central Europe.

For the case of economically optimal line volumes, there is a strong
transmission grid expansion over all of Europe that allows to balance synoptic-scale weather variations and for optimal utilization
of the best resource locations, making the trends discussed above more pronounced. This large grid allows significant net exports of onshore wind
generation from a few countries like Denmark, Great Britain, Ireland, Norway, and Sweden. The first two of these are located slightly more centrally in the grid and require less H$_2$ storage than in the previous scenario, while the latter three are further away from the importers, and build more H$_2$ storage.

The solar generation is concentrated in southern Europe where the highest solar full load hours are
found, but still require net imports. The South-East has the smallest grid expansion, indicating that
solar generation does not directly profit as much from (local) spatial smoothing
effects.
Greece plays an important role in this region
to diversify the power sources with its high share of offshore wind that can be
exported, e.g., during the night.

\subsection{Marginal prices}

The average marginal prices for each node derived from the KKT multipliers $\lambda_{n,t}$ in equation \eqref{eq:balance} are plotted in
Fig.~\ref{fig:lmp} for each node and for three scenarios. Here several trends are noticeable. First, the overall prices decrease as transmission is increased, reflecting the decrease in total system costs. The spread of prices between the nodes also narrows as increased transmission reduces congestion, allowing prices to equalise within the network. Finally we see that the overall prices are slightly higher than the average system costs, because they include the scarcity costs induced by the constraints, such as the CO$_2$ cost which contributes on average 5~\euro/MWh and the scarcity costs of generation sites with limited expansion potential.

The lowest prices are in those countries with high shares of zero-marginal-cost renewable generation, such as Denmark, Ireland, Sweden and Norway. Norway actually sees an increase in prices as transmission is expanded, since it can share its abundant hydro resources with other countries, to the benefit of the entire system.

\subsection{Line volume shadow price}

The economic value of transmission line volumes can be analyzed with the help of the shadow price $\mu_{LV}$ of the line volume constraint defined in eq.~\eqref{eq:lvcap}. The shadow price is the dual variable of this constraint and can be interpreted as the price per unit of line volume the system is willing to pay to build a given amount $LV$ of constrained line volume. In other words, this is the cost required so that in an unconstrained market, the economic optimum is to build a total line volume of $LV$.

\begin{figure}%
\centering
\includegraphics[width=\linewidth]{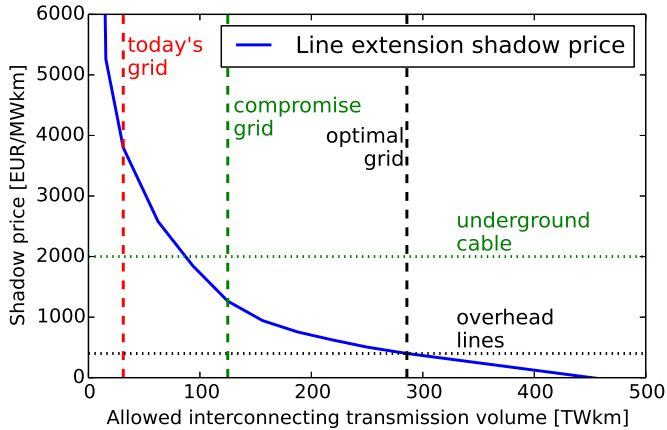}%
\caption{Shadow price of the line volume constraint in units of overnight capital costs as function of the allowed total line volume. As in Fig.~\ref{fig:costs_LV}, the dashed vertical lines mark the transmission line volumes of today's grid (red), the compromise grid (green) at four times today's volume, and the economically optimal grid (black). The dotted horizontal lines indicate the capital costs of overhead transmission lines (black) and underground cables (green).
}
\label{fig:shadow_price_LV}
\end{figure}

Therefore, at the assumed current costs for over-land HVDC transmission lines of 400~\euro/MWkm, the model finds the optimal grid volume of 286 TWkm, as marked by the black lines in Fig.~\ref{fig:shadow_price_LV}. Here, the plotted shadow price is defined such that it represents the overnight capital costs of lines.
This figure allows to read off the points at
which more expensive transmission solutions, such as underground
cabling, make economic sense.
If all lines would be replaced by underground cables with a current cost of ca. 2000~\euro/MWkm, a grid extension to 90 TWkm, roughly three times the current NTC would still be economically optimal in this model.
The proposed compromise grid with four times today's volumes derived from considerations of the total system costs would still be built if transmission would cost 1300 \euro/MWkm.
In a purely market based solution, transmission would have to cost 4000~\euro/MWkm in order to limit the optimal line volume to today's NTC in this model.

\subsection{Dispatch time series}
\begin{figure}%
\centering
\includegraphics[width=\linewidth]{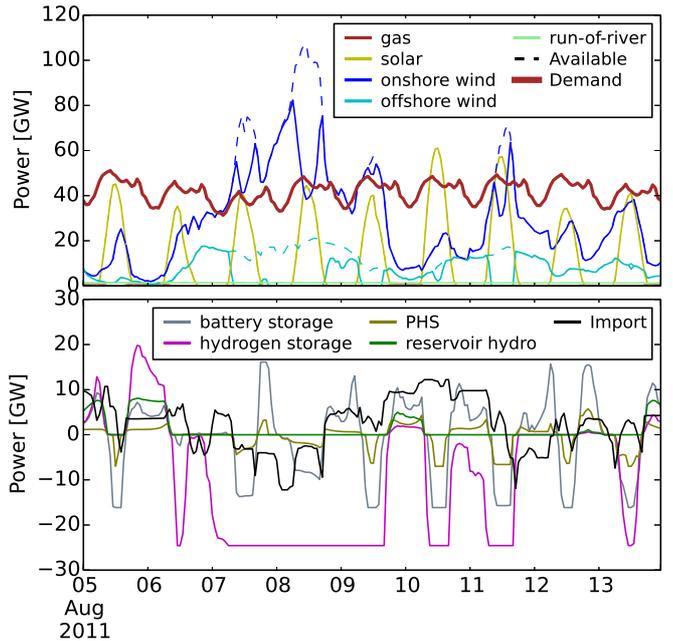}
\caption{Example time series of the optimized hourly dispatch of the different technologies in France based on historic demand and weather data from August 5th to 14th 2011 in the scenario with today's interconnecting transmission volumes. The actual and available hourly power outputs of generators are respectively shown as solid and dashed lines. Discharging of storage units is indicated by positive, charging by negative values. Positive values of the black solid line mark imports into France, negative values mark exports. The thick brown line shows the demand. Note the different y-scale for top and bottom panel.
}
\label{fig:dispatch}
\end{figure}

In this subsection some example dispatch time series from the model are examined. Fig.~\ref{fig:dispatch} shows an example from France in August with high generation from both wind and solar. Some onshore and offshore wind has to be curtailed, as shown by the difference between available power (dashed lines) and dispatched power (solid lines). More offshore than onshore wind is curtailed, but no solar generation, following the assumed curtailment ordering via marginal costs as described above.

During this period, hydrogen storage is charged at full power most of the time when there is excess wind generation, and is only discharged in a few nights with very little wind generation. Similarly, reservoir hydro is also only dispatched during these nights and accumulates the inflow the rest of the time, which is not shown here.
Batteries and pumped hydro storage (PHS) provide peak shaving between day and night, where they are charged mostly from solar during the day and discharge during the night.
The imports and exports tend to be correlated with local under- and overproduction of wind, respectively, but they also depend strongly on the state of the rest of the network.
The dispatch of run-of-river is almost constant and very small, and gas power is not generated during this time.

\begin{figure}%
\centering
\includegraphics[width=\linewidth]{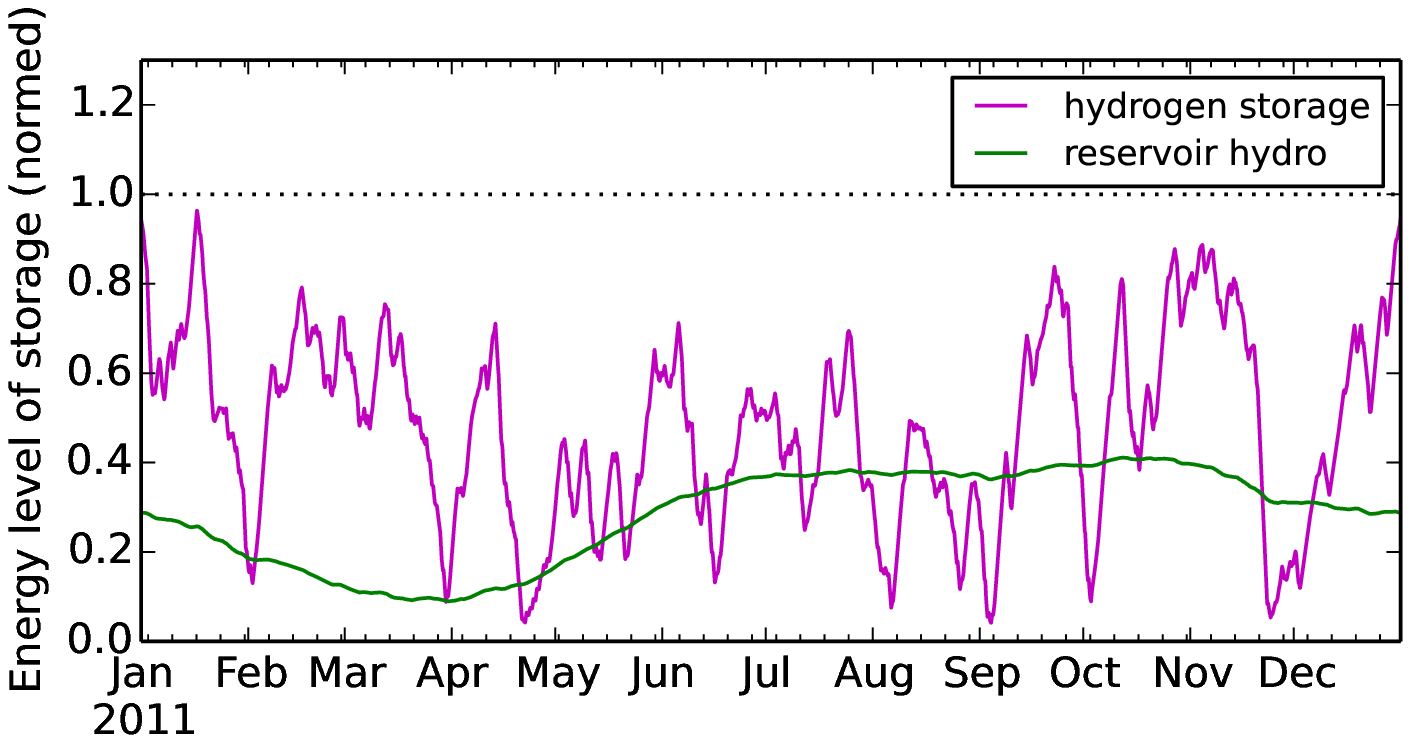}
\includegraphics[width=\linewidth]{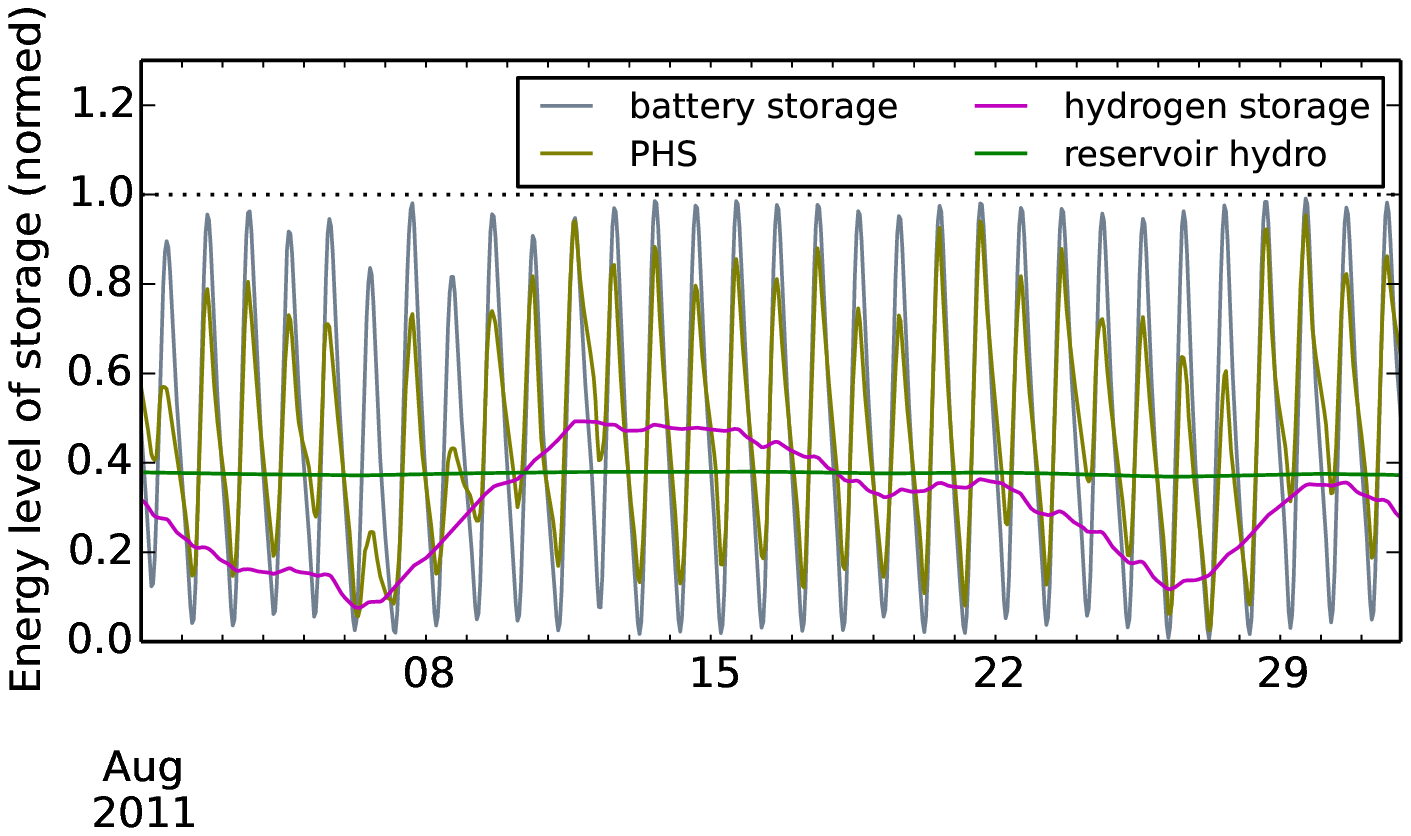}
\caption{Hourly total energy storage levels for the compromise grid scenario, normalized to the storage energy capacity. Battery storage and PHS levels show a similar, predominantly daily pattern, leading to some visual overlap of the lines.
}
\label{fig:soc}
\end{figure}

Fig.~\ref{fig:soc} shows the behaviour of the state of charge for all the storage in Europe, revealing the different temporal scales on which each technology operates. Reservoir hydro shows a seasonal pattern, discharging in winter when demand is high, and charging in spring and summer as snow melts in mountainous areas. The sum of reservoir storage levels never drops to zero because it is aggregated over several countries; individual countries do drop to zero, but at different times. Hydrogen storage varies on seasonal and synoptic scales, reflecting the pairing of this long-term storage with wind. Finally, battery and PHS show a daily pattern reflecting the use of this storage resource to balance variations in solar generation.

\section{Discussion}
\label{sec:diss}

\subsection{Comparison to the literature}

Average system costs range in this study from 64.8~\euro{}/MWh in the case of optimal transmission up to 84.1~\euro/MWh with no inter-connecting transmission. These values are consistent with other values found in the literature. For example, Czisch \cite{Czisch} used similar cost assumptions (with the exception of the then-reasonable overnight cost of 5500~\euro/kW for PV) and a target of 100\% CO$_2$-free generation and found average costs of 46.5~\euro/MWh for optimal transmission in a system comprising Europe, the Middle East and North Africa (EUMENA), 52~\euro/MWh with no transmission between Africa and Europe, and 80~\euro/MWh with no transmission at all between countries. Scholz \cite{Scholz} found for a 100\% renewable system for EUNA an average cost of 69~\euro/MWh with optimal transmission and 83~\euro/MWh with no transmission.
Bussar et al. \cite{Bussar201440} found for EUMENA an optimal cost of 69~\euro/MWh and a total line volume of 375~TWkm with optimal transmission. These costs compare well to the extreme points of our analysis. Analysing the full spectrum of possible network extensions between the extreme points, as has been done here, reveals the non-linear development of the costs and the benefits of a compromise transmission expansion.

In \cite{rolando2014,sarah} similar non-linear effects of transmission were found in a more simplified model of a highly renewable European electricity system without a limit on CO$_2$ emissions, without storage or hydroelectricity, and without optimising transmission and generator capacities. In that model, if there were no  constraints on transmission, the model would build 12 times today's inter-connecting capacities to minimise the need for backup energy; however, 90\% of the benefit was found with 4 times today's capacities and 70\% of the benefit with 2 times today's capacities. In \cite{OptHet} this approach was extended to include optimisation of wind and solar capacity placement, but again the results are not comparable because CO$_2$ emissions were not limited (which would translate into a limit on backup energy from fossil fuels), and storage and hydroelectricity were not considered. However, the average system cost range found in \cite{OptHet} of 53~\euro/MWh to 64.5~\euro/MWh, depending on the amount of transmission and the level of optimisation of generation capacities, is also comparable to the studies mentioned above.

Next we turn to studies that optimise grid capacity in Europe with
more network nodes, so that transmission lines inside each country are
also seen in the model. These models tend to see a lower overall level
of grid expansion at the cost-optimal level than studies which
aggregate each country to a single node. For example, in
\cite{Hagspiel} with a 200-node model of Europe, the cost-optimal grid capacity for a 90\% CO$_2$
reduction in the electricity sector is around double today's total
capacity. Similar results can be found in \cite{Schaber,Schaber2} with up to 83 nodes. Lower levels of grid expansion are
seen when looking at lower levels of renewables in the near future
\cite{Brown,TYNDP2016}. The dependence of electricity system optimisation results on the level of spatial resolution (i.e. the number of network nodes) was studied in detail recently in \cite{hoersch2017arxiv}, where it was also found that the cost-optimal level of grid expansion for a 95\% CO$_2$ reduction was around 2-3 times today's capacities when including country-internal transmission lines; using more expensive underground cables for the grid expansion results in optimal networks that are only 30\% to 60\% bigger than today. Models that only include transmission lines between countries, like the one presented in this paper, show bigger network expansions because interconnectors have traditionally been weaker than the networks inside countries, given that the current interconnected system has evolved slowly by combining national systems.

In common with all the studies mentioned above, the model presented here shows the dominance of wind power (up to 65\% of total energy production) when transmission expansion is allowed.

\subsection{Limitations of the study}

As discussed above, one limitation of this study is that grid
bottlenecks inside each country are not considered in the model, since
we chose to focus on the benefits of international transmission. It
was shown in \cite{hoersch2017arxiv} that while expanding the grid to
resolve these country-internal bottlenecks might double transmission
costs in the cost-optimal case, these costs are still less than 10\%
of total system costs. If national networks cannot be expanded because of
acceptance problems, these bottlenecks cause additional costs by
restricting generation feed-in, forcing a shift from offshore wind
expansion to more local production from solar and storage.

In this study the costs of possible expansion of the distribution grid
and the provision of ancillary services have not been
considered. Studies which have considered the expansion of the
distribution grid for high shares of renewables show that these
costs range from around 10\% to 15\% of total system costs
\cite{DENAII,ICNERA,RLP}; taking account of ancillary services, such as
voltage control, frequency control, fault current provision and
black-start capabilities have also been shown to play only a secondary
role in the total system costs \cite{KWK2,DENASS,urdal2015}.

Another limitation of this study is that synergies arising from
coupling the electricity sector to other energy sectors, such as
transport and heating, have been neglected.  A follow-up study to this
one is being prepared that considers sector-coupling in a European
context.  Preliminary results \cite{Brown2016} show that the
coordinated charging of battery electric vehicles can replace the role
of stationary batteries in balancing solar variations, while the
longer-term variations of wind can be accommodated using Power-To-Gas
and Power-To-Heat units in combination with thermal energy
storage. This sector coupling can further reduce the need for
inter-connecting transmission.

\section{Summary and Conclusions}
\label{sec:concl}

In this paper a techno-economic model was implemented and optimised to examine the effect of different levels of inter-connecting transmission on the costs of the European electricity system, assuming a reduction of CO$_2$ emissions of 95\% compared to 1990 levels.
This model includes renewable energy generation from wind, solar and hydro power, and storage systems such as pumped hydro storage, batteries, and hydrogen storage units.

We interpolate continuously between a cost-optimized level of inter-connecting transmission and no or limited inter-connection between European countries. We analyse the average system costs depending on different transmission volume levels.
This reveals non-linear effects which are
not visible in  studies of isolated transmission scenarios \cite{Czisch,Scholz,Bussar201440}; in particular, most of the
economic benefits of transmission expansion can be locked in with a more modest
expansion than the optimal solution. An expansion to four times today's inter-connection capacities already enables 85\% of the cost savings of the optimal transmission expansion (9 times today's).  This conclusion has important
policy consequences, because it offers a compromise between the needs
of the electricity system and low acceptance of transmission grid expansion by the public.

With the cost-optimal level of transmission, the European electricity
system can be built with a total average system cost as low as
64.8~\euro{}/MWh, comparable to the cost of the current system without CO$_2$ pricing. This
system uses a continent-wide transmission grid to balance the
large-scale synoptic variations of wind (65\% of energy generation)
and integrate hydroelectricity from mountainous regions (15\% of
generation). Restricting transmission drives total costs up by a third, because the grid is no longer available to balance the
variations of wind in space. Instead, long-term hydrogen storage must be used
to balance the variations of wind over several days; since this is
comparatively expensive, restricting transmission favours a
combination of solar generation (up to 36\%) with daily battery
storage. This shows the importance of considering spatial and temporal
scales when analysing the integration of renewables. It also shows
that there is no single solution for a highly renewable electricity
system. Instead there is a family of possible solutions with different
properties (such as level of transmission) and different costs.

This study has highlighted the importance and cost-efficiency of a global European energy transmission network, even if it can not be extended to its economically optimal size due to external limitations like public acceptance issues. The flatness of the costs around the optimal point as transmission is restricted is a general feature that also applies to other types of restrictions.
In a forthcoming study, this work will be extended to the exploration
of other directions that also turn out to be flat in the optimization space, such as
restrictions on the import levels of each country, restrictions on onshore wind due to public acceptance problems, and variations of the CO$_2$ cap or price.
These flat directions are important, because they may allow solutions that are
both cost-effective and also take account of other political restrictions. In a further study, the benefits of coupling to other 
energy sectors, such as
transport and heating will be considered.

These studies will lead to a much better understanding of the importance of European cooperation in terms of energy distribution, and it will contribute to a more stable, cost-efficient and effective setting for the future stability of a highly renewable European energy network.

\section*{Acknowledgements}

The authors thank %
the anonymous reviewers for their helpful suggestions and comments that improved this work.
The project underlying this report was supported by the German Federal Ministry
of Education and Research under grant no.~03SF0472C. The responsibility for the
contents lies with the authors.

\bibliographystyle{elsarticle-num}
\biboptions{sort}
\bibliography{literatur}

\begin{thebibliography}{10}
\expandafter\ifx\csname url\endcsname\relax
  \def\url#1{\texttt{#1}}\fi
\expandafter\ifx\csname urlprefix\endcsname\relax\def\urlprefix{URL }\fi
\expandafter\ifx\csname href\endcsname\relax
  \def\href#1#2{#2} \def\path#1{#1}\fi

\bibitem{roadm2050}
{EC}, {Energy roadmap 2050 -- COM(2011) 885/2} (2011).

\bibitem{Czisch}
G.~Czisch, Szenarien zur zuk\"unftigen {S}tromversorgung, Ph.D. thesis,
  Universit\"at Kassel (2005).

\bibitem{Scholz}
Y.~Scholz, {Renewable energy based electricity supply at low costs -
  Development of the REMix model and application for Europe}, Ph.D. thesis,
  Universit\"at Stuttgart (2012).
\newblock \href {http://dx.doi.org/10.18419/opus-2015}
  {\path{doi:10.18419/opus-2015}}.

\bibitem{Wagner2016}
F.~Wagner, Surplus from and storage of electricity generated by intermittent
  sources, The European Physical Journal Plus 131~(12) (2016) 445.
\newblock \href {http://dx.doi.org/10.1140/epjp/i2016-16445-3}
  {\path{doi:10.1140/epjp/i2016-16445-3}}.

\bibitem{Schaber}
K.~Schaber, F.~Steinke, T.~Hamacher, Transmission grid extensions for the
  integration of variable renewable energies in {E}urope: Who benefits where?,
  Energy Policy 43 (2012) 123 -- 135.
\newblock \href {http://dx.doi.org/10.1016/j.enpol.2011.12.040}
  {\path{doi:10.1016/j.enpol.2011.12.040}}.

\bibitem{Schaber2}
K.~Schaber, F.~Steinke, P.~M{\"u}hlich, T.~Hamacher, Parametric study of
  variable renewable energy integration in {E}urope: Advantages and costs of
  transmission grid extensions, Energy Policy 42 (2012) 498--508.
\newblock \href {http://dx.doi.org/10.1016/j.enpol.2011.12.016}
  {\path{doi:10.1016/j.enpol.2011.12.016}}.

\bibitem{rolando2014}
R.~A. {Rodriguez}, S.~{Becker}, G.~B. {Andresen}, D.~{Heide}, M.~{Greiner},
  {Transmission needs across a fully renewable {E}uropean power system},
  Renewable Energy 63 (2014) 467--476.

\bibitem{Hagspiel}
S.~Hagspiel, C.~J{\"a}gemann, D.~Lindenberger, T.~Brown, S.~Cherevatskiy,
  E.~Tr{\"o}ster, Cost-optimal power system extension under flow-based market
  coupling, Energy 66 (2014) 654--666.
\newblock \href {http://dx.doi.org/10.1016/j.energy.2014.01.025}
  {\path{doi:10.1016/j.energy.2014.01.025}}.

\bibitem{Brown}
T.~Brown, P.-P. Schierhorn, E.~Tr{\"o}ster, T.~Ackermann, Optimising the
  {European} transmission system for 77\% renewable electricity by 2030, IET
  Renewable Power Generation 10~(1) (2016) 3--9.
\newblock \href {http://dx.doi.org/10.1049/iet-rpg.2015.0135}
  {\path{doi:10.1049/iet-rpg.2015.0135}}.

\bibitem{Battaglinietal2012}
A.~Battaglini, N.~Komendantova, P.~Brtnik, A.~Patt, Perception of barriers for
  expansion of electricity grids in the {E}uropean {U}nion, Energy Policy 47
  (2012) 254--259.
\newblock \href {http://dx.doi.org/10.1016/j.enpol.2012.04.065}
  {\path{doi:10.1016/j.enpol.2012.04.065}}.

\bibitem{Heide2010}
D.~Heide, L.~von Bremen, M.~Greiner, C.~Hoffmann, M.~Speckmann, S.~Bofinger,
  Seasonal optimal mix of wind and solar power in a future, highly renewable
  {Europe}, Renewable {En}ergy 35~(11) (2010) 2483--2489.
\newblock \href {http://dx.doi.org/10.1016/j.renene.2010.03.012}
  {\path{doi:10.1016/j.renene.2010.03.012}}.

\bibitem{sarah}
S.~{Becker}, R.~A. {Rodr\'iguez}, G.~B. {Andresen}, S.~{Schramm}, M.~{Greiner},
  {Transmission grid extensions during the build-up of a fully renewable
  pan-{E}uropean electricity supply}, Energy 64 (2014) 404--418.

\bibitem{Sensitivity}
R.~A. Rodriguez, S.~Becker, M.~Greiner, Cost-optimal design of a simplified,
  highly renewable pan-{E}uropean electricity system, Energy 83 (2015) 658 --
  668.
\newblock \href {http://dx.doi.org/10.1016/j.energy.2015.02.066}
  {\path{doi:10.1016/j.energy.2015.02.066}}.

\bibitem{OptHet}
E.~H. Eriksen, L.~J. Schwenk-Nebbe, B.~Tranberg, T.~Brown, M.~Greiner, Optimal
  heterogeneity in a simplified highly renewable {E}uropean electricity system,
  Energy (2017) --\href {http://dx.doi.org/10.1016/j.energy.2017.05.170}
  {\path{doi:10.1016/j.energy.2017.05.170}}.

\bibitem{Bussar201440}
C.~Bussar, M.~Moos, R.~Alvarez, P.~Wolf, T.~Thien, H.~Chen, Z.~Cai,
  M.~Leuthold, D.~U. Sauer, A.~Moser, Optimal allocation and capacity of energy
  storage systems in a future {E}uropean power system with 100\% renewable
  energy generation, Energy Procedia 46 (2014) 40 -- 47, 8th International
  Renewable Energy Storage Conference and Exhibition (IRES 2013).
\newblock \href {http://dx.doi.org/10.1016/j.egypro.2014.01.156}
  {\path{doi:10.1016/j.egypro.2014.01.156}}.

\bibitem{Egerer}
J.~Egerer, C.~Lorenz, C.~Gerbaulet, {E}uropean electricity grid infrastructure
  expansion in a 2050 context, in: 10th International Conference on the
  {E}uropean Energy Market, IEEE, 2013, pp. 1--7.
\newblock \href {http://dx.doi.org/10.1109/EEM.2013.6607408}
  {\path{doi:10.1109/EEM.2013.6607408}}.

\bibitem{mathiesen2014smart}
B.~V. Mathiesen, H.~Lund, D.~Conolly, H.~Wenzel, P.~{\O}stergaard,
  B.~M{\"o}ller, S.~Nielsen, I.~Ridjan, P.~Karn{\o}e, K.~Sperling,
  F.~Hvelplund, Smart energy systems for coherent 100\% renewable energy and
  transport solutions, Applied Energy 145 (2015) 139--154.
\newblock \href {http://dx.doi.org/10.1016/j.apenergy.2015.01.075}
  {\path{doi:10.1016/j.apenergy.2015.01.075}}.

\bibitem{Lund201296}
H.~Lund, A.~N. Andersen, P.~A. Østergaard, B.~V. Mathiesen, D.~Connolly, From
  electricity smart grids to smart energy systems – a market operation based
  approach and understanding, Energy 42~(1) (2012) 96 -- 102, 8th World Energy
  System Conference, {WESC} 2010.
\newblock \href {http://dx.doi.org/10.1016/j.energy.2012.04.003}
  {\path{doi:10.1016/j.energy.2012.04.003}}.

\bibitem{Henning20141003}
H.-M. Henning, A.~Palzer, {A comprehensive model for the German electricity and
  heat sector in a future energy system with a dominant contribution from
  renewable energy technologies—Part I: Methodology}, Renewable and
  Sustainable Energy Reviews 30 (2014) 1003 -- 1018.
\newblock \href {http://dx.doi.org/10.1016/j.rser.2013.09.012}
  {\path{doi:10.1016/j.rser.2013.09.012}}.

\bibitem{IEESWV}
N.~Gerhardt, A.~Scholz, F.~Sandau, H.~Hahn, {Interaktion EE-Strom, Wärme und
  Verkehr}, Tech. rep., Fraunhofer IWES,
  \url{http://www.energiesystemtechnik.iwes.fraunhofer.de/de/projekte/suche/2015/interaktion_strom_waerme_verkehr.html}
  (2015).

\bibitem{Quaschning}
V.~Quaschning, {Sektorkopplung durch die Energiewende}, Tech. rep., HTW Berlin
  (2016).

\bibitem{Deane2012303}
J.~Deane, A.~Chiodi, M.~Gargiulo, B.~P.~O. Gallachoir, Soft-linking of a power
  systems model to an energy systems model, Energy 42~(1) (2012) 303 -- 312,
  8th World Energy System Conference, \{WESC\} 2010.
\newblock \href {http://dx.doi.org/10.1016/j.energy.2012.03.052}
  {\path{doi:10.1016/j.energy.2012.03.052}}.

\bibitem{Connolly20161634}
D.~Connolly, H.~Lund, B.~Mathiesen, {Smart Energy Europe: The technical and
  economic impact of one potential 100\% renewable energy scenario for the
  European Union}, Renewable and Sustainable Energy Reviews 60 (2016) 1634 --
  1653.
\newblock \href {http://dx.doi.org/10.1016/j.rser.2016.02.025}
  {\path{doi:10.1016/j.rser.2016.02.025}}.

\bibitem{Brown2016}
T.~Brown, D.~Schlachtberger, A.~Kies, M.~Greiner, Sector coupling in a
  simplified model of a highly renewable {E}uropean energy system, in:
  Proceedings of 15th Wind Integration Workshop, 2016.

\bibitem{PyPSA}
T.~Brown, J.~H{\"o}rsch, D.~Schlachtberger, {Python for Power System Analysis
  (PyPSA) Version 0.9.0}, 2017.
\newblock \href {http://dx.doi.org/10.5281/zenodo.582307}
  {\path{doi:10.5281/zenodo.582307}}.

\bibitem{zenodo}
D.~Schlachtberger, T.~Brown, S.~Schramm, M.~Greiner, {Supplementary Data: The
  Benefits of Cooperation in a Highly Renewable European Electricity Network}
  (Jun. 2017).
\newblock \href {http://dx.doi.org/10.5281/zenodo.804338}
  {\path{doi:10.5281/zenodo.804338}}.

\bibitem{Schweppeetal1988}
F.~C. Schweppe, M.~C. Caramanis, R.~D. Tabors, R.~E. Bohn, Spot Pricing of
  Electricity, Norwell, MA: Kluwer, 1988.

\bibitem{Biggar}
D.~R. Biggar, M.~R. Hesamzadeh, The Economics of Electricity Markets, Wiley,
  2014.

\bibitem{DENAII}
{Deutsche Energie-Agentur}, {DENA-Netzstudie II}, online at
  \url{http://www.dena.de/publikationen/energiesysteme} (2010).

\bibitem{TYNDP2016}
{European Network of Transmission System Operators for Electricity}, {Ten-Year
  Network Development Plan (TYNDP) 2016}, Tech. rep., ENTSO-E (2016).

\bibitem{hoersch2017arxiv}
{H\"orsch, J.}, {Brown, T.}, The role of spatial scale in joint optimisations
  of generation and transmission for {E}uropean highly renewable scenarios, in:
  {Proceedings of 14th International Conference on the European Energy Market
  (EEM 2017)}, 2017.

\bibitem{gurobi}
{Gurobi Optimization, Inc.}, Gurobi optimizer reference manual,
  \url{http://www.gurobi.com} (2016).

\bibitem{entsoe_load}
{European Transmission System Operators}, {Country-specific hourly load data},
  \url{https://www.entsoe.eu/data/data-portal/consumption/} (2011).

\bibitem{Heide2011}
D.~Heide, M.~Greiner, L.~Von~Bremen, C.~Hoffmann, Reduced storage and balancing
  needs in a fully renewable {E}uropean power system with excess wind and solar
  power generation, Renewable Energy 36~(9) (2011) 2515--2523.
\newblock \href {http://dx.doi.org/10.1016/j.renene.2011.02.009}
  {\path{doi:10.1016/j.renene.2011.02.009}}.

\bibitem{Andresen20151074}
G.~B. Andresen, A.~A. S{\o}ndergaard, M.~Greiner, Validation of {D}anish wind
  time series from a new global renewable energy atlas for energy system
  analysis, Energy 93, Part 1 (2015) 1074 -- 1088.
\newblock \href {http://dx.doi.org/10.1016/j.energy.2015.09.071}
  {\path{doi:10.1016/j.energy.2015.09.071}}.

\bibitem{corine2006}
EEA, Corine land cover 2006 (2014).

\bibitem{natura2000}
EEA, Natura 2000 data - the {E}uropean network of protected sites,
  \url{http://www.eea.europa.eu/data-and-maps/data/natura-7} (2016).

\bibitem{kies2016}
A.~Kies, K.~Chattopadhyay, L.~von Bremen, E.~Lorenz, D.~Heinemann, {RESTORE
  2050 Work Package Report D12: Simulation of renewable feed-in for power
  system studies.}, Tech. rep., RESTORE 2050, in preparation (2016).

\bibitem{pfluger2011}
B.~Pfluger, F.~Sensfu{\ss}, G.~Schubert, J.~Leisentritt, {Tangible ways towards
  climate protection in the European Union (EU Long-term scenarios 2050)},
  Fraunhofer ISI.

\bibitem{ENTSOEinstalledcapas}
{European Transmission System Operators}, {Installed Capacity per Production
  Type in 2015}, Tech. rep., ENTSO-E (2016).

\bibitem{dee2011}
D.~P. Dee, S.~M. Uppala, A.~J. Simmons, P.~Berrisford, P.~Poli, S.~Kobayashi,
  U.~Andrae, M.~A. Balmaseda, G.~Balsamo, P.~Bauer, P.~Bechtold, A.~C.~M.
  Beljaars, L.~van~de Berg, J.~Bidlot, N.~Bormann, C.~Delsol, R.~Dragani,
  M.~Fuentes, A.~J. Geer, L.~Haimberger, S.~B. Healy, H.~Hersbach, E.~V. Hólm,
  L.~Isaksen, P.~Kållberg, M.~Köhler, M.~Matricardi, A.~P. McNally, B.~M.
  Monge-Sanz, J.-J. Morcrette, B.-K. Park, C.~Peubey, P.~de~Rosnay,
  C.~Tavolato, J.-N. Thépaut, F.~Vitart, The {ERA-Interim} reanalysis:
  configuration and performance of the data assimilation system, Quarterly
  Journal of the Royal Meteorological Society 137~(656) (2011) 553--597.
\newblock \href {http://dx.doi.org/10.1002/qj.828} {\path{doi:10.1002/qj.828}}.

\bibitem{schroeder2013}
A.~Schr\"{o}der, F.~Kunz, J.~Meiss, R.~Mendelevitch, C.~von Hirschhausen,
  Current and prospective costs of electricity generation until 2050, Data
  Documentation, DIW~68, Deutsches Institut f\"{u}r Wirtschaftsforschung (DIW),
  Berlin, \url{http://hdl.handle.net/10419/80348}, accessed July 2016 (2013).

\bibitem{budischak2013}
C.~Budischak, D.~Sewell, H.~Thomson, L.~Mach, D.~E. Veron, W.~Kempton,
  Cost-minimized combinations of wind power, solar power and electrochemical
  storage, powering the grid up to 99.9\% of the time, Journal of Power Sources
  225 (2013) 60 -- 74.
\newblock \href {http://dx.doi.org/10.1016/j.jpowsour.2012.09.054}
  {\path{doi:10.1016/j.jpowsour.2012.09.054}}.

\bibitem{ENTSOEYSAR2013}
{European Transmission System Operators}, {Yearly Statistics \& Adequacy
  Retrospect 2013 background data},
  \url{https://www.entsoe.eu/Documents/Publications/Statistics/YSAR/160112_YSAR_2013_data.zip}
  (2016).

\bibitem{ICNERA}
{Integration of Renewable Energy in Europe}, Tech. rep., Imperial College,
  NERA, DNV GL,
  \url{https://ec.europa.eu/energy/sites/ener/files/documents/201406_report_renewables_integration_europe.pdf}
  (2014).

\bibitem{RLP}
T.~Ackermann, S.~Untsch, M.~Koch, H.~Rothfuchs, {Verteilnetzstudie
  Rheinland-Pfalz}, 2014,
  \url{https://www.oeko.de/oekodoc/1885/2014-008-de.pdf}.

\bibitem{KWK2}
K.~Knorr, et~al., Kombikraftwerk 2: Abschlussbericht, Tech. rep., Fraunhofer
  IWES and others (August 2014).

\bibitem{DENASS}
{dena-Studie Systemdienstleistungen 2030}, Tech. rep., DENA, online at
  \url{https://www.dena.de/themen-projekte/projekte/energiesysteme/dena-studie-systemdienstleistungen-2030/}
  (2014).

\bibitem{urdal2015}
H.~Urdal, R.~Ierna, J.~Zhu, C.~Ivanov, A.~Dahresobh, D.~Rostom, System strength
  considerations in a converter dominated power system, IET Renewable Power
  Generation 9 (2015) 10--17(7).
\newblock \href {http://dx.doi.org/10.1049/iet-rpg.2014.0199}
  {\path{doi:10.1049/iet-rpg.2014.0199}}.

\end{thebibliography}

\end{document}